\documentclass[aps,prc,superscriptaddress,reprint]{revtex4-2}
\usepackage{graphicx}
\usepackage{multirow}
\usepackage{color}
\usepackage{amssymb}
\usepackage{bbm}
\usepackage{amsmath}
\usepackage{lineno}
\usepackage{enumitem}
\usepackage{scalerel,stackengine}
\usepackage{xcolor}
\usepackage{comment}
\usepackage{booktabs}
\usepackage[normalem]{ulem}
\usepackage[colorlinks=true, breaklinks=true, linkcolor=blue, citecolor=purple, urlcolor=teal]{hyperref}

\renewcommand{\Im}{\mathop{\mathrm{Im}}}
\renewcommand{\Re}{\mathop{\mathrm{Re}}}

\def\nuc#1#2{\relax\ifmmode{}^{#1}{\protect\text{#2}}\else${}^{#1}$#2\fi}

\newcommand{\be}{\begin{eqnarray}}
\newcommand{\ee}{\end{eqnarray}}

\newcommand{\bwt}{\begin{widetext}}
\newcommand{\ewt}{\end{widetext}}

\stackMath
\newcommand\reallywidehat[1]{%
\savestack{\tmpbox}{\stretchto{%
  \scaleto{%
    \scalerel*[\widthof{\ensuremath{#1}}]{\kern-.6pt\bigwedge\kern-.6pt}%
    {\rule[-\textheight/2]{1ex}{\textheight}}%
  }{\textheight}%
}{0.5ex}}%
\stackon[1pt]{#1}{\tmpbox}%
}

\begin{document}

\title{Direct Boundary Matching: A Bound-State Technique for Nuclear Scattering with Lagrange-Legendre Functions}

\author{Jin Lei}
\email[]{jinl@tongji.edu.cn}
\affiliation{School of Physics Science and Engineering, Tongji University, Shanghai 200092, China.}

\begin{abstract}
I present a direct boundary matching method (DBMM) for solving nuclear scattering problems using Lagrange-Legendre basis functions. This approach belongs to the family of bound-state techniques for the continuum, reformulating scattering problems into a localized, square-integrable ($L^2$) representation. The key feature is the direct incorporation of the outgoing wave boundary condition into the last row of the matrix equation, eliminating the need for Bloch operators and two-step matching procedures required in traditional R-matrix methods. Unlike the complex scaling method that rotates coordinates into the complex plane, DBMM operates entirely in real coordinate space. The formalism is extended to coupled-channel problems, where the wave function decomposition naturally leads to an effective source potential that distinguishes between the entrance channel and other channels. Benchmark calculations for p~+~$^{12}$C scattering demonstrate excellent agreement with the Numerov integration method.
\end{abstract}


\date{\today}

\maketitle

\section{Introduction}
\label{sec:intro}

The theoretical description of quantum scattering presents fundamental challenges that distinguish it from bound-state calculations. While bound-state wave functions decay exponentially at large distances, allowing them to be accurately represented within finite computational domains, scattering states exhibit oscillatory asymptotic behavior that extends to infinity~\cite{Taylor1972}. This fundamental difference has historically necessitated specialized methods for continuum problems, creating a dichotomy between nuclear structure and nuclear reaction theory.

In bound-state calculations, the radial wave function $\psi(r)$ satisfies
\begin{equation}
\psi(r) \xrightarrow{r \to \infty} e^{-\kappa r},
\end{equation}
where $r$ is the radial coordinate, $\kappa = \sqrt{2\mu|E|}/\hbar$ characterizes the exponential decay, $\mu$ is the reduced mass of the two-body system, and $E < 0$ is the binding energy. This localization permits accurate approximation using truncated bases of square-integrable ($L^2$) functions, such as harmonic oscillator states or Gaussian expansions. Standard diagonalization techniques then yield the discrete energy spectrum and associated eigenfunctions.

Scattering states, in contrast, exist at positive energies forming a continuous spectrum. The asymptotic wave function for a partial wave $\ell$ takes the form
\begin{equation}
\psi_\ell(r) \xrightarrow{r \to \infty} F_\ell(\eta, kr) + k f_\ell H_\ell^+(\eta, kr),
\end{equation}
where $F_\ell$ and $H_\ell^+ = G_\ell + iF_\ell$ denote the regular Coulomb function and outgoing Coulomb-Hankel function (with $G_\ell$ being the irregular Coulomb function), respectively, $\eta$ is the Sommerfeld parameter, and $f_\ell$ is the scattering amplitude to be determined. The oscillatory nature of these functions precludes straightforward $L^2$ representation, as any truncated basis of decaying functions cannot capture the infinite-range oscillations.

Traditional approaches to this problem involve numerical integration of the radial Schr\"{o}dinger equation followed by asymptotic matching. For coupled-channel problems, one must solve systems of coupled differential equations, which becomes increasingly cumbersome as the number of channels grows.

In response to these challenges, a class of methods known collectively as ``bound-state techniques for the continuum'' has emerged over the past decades~\cite{Carbonell2014,Johnson2020}. The unifying philosophy is to reformulate the scattering problem, inherently defined over an infinite domain with oscillatory asymptotics, into a form solvable within a localized, $L^2$ basis. This approach leverages the computational machinery developed for nuclear structure to access continuum observables.

Several distinct realizations of this philosophy have been developed. The \textit{R-matrix method}~\cite{Wigner1947,Descouvemont2010,Jin20} divides configuration space into internal and external regions, using a Bloch operator to ensure matrix symmetry in the internal region; scattering amplitudes are then extracted through a two-step matching procedure. The \textit{Complex Scaling Method} (CSM)~\cite{Kruppa2007,Lazauskas2011,Liu2025,Moiseyev1998,Myo2014} rotates coordinates into the complex plane ($r \to r e^{i\theta}$), converting oscillatory scattering states into $L^2$-integrable functions. The \textit{Lorentz Integral Transform} (LIT) method~\cite{Efros1994,Efros2007} employs a Lorentzian kernel to reduce continuum response functions to bound-state-like calculations, with physical observables recovered through numerical inversion. The \textit{Kohn variational principle}~\cite{Kievsky1997,Viviani2020} directly yields scattering amplitudes through a variational functional that incorporates correct asymptotic behavior in the trial wave function.

In this work, I present a direct boundary matching method (DBMM) that offers an alternative realization of the bound-state technique philosophy. It should be emphasized that the novelty of DBMM lies not in the individual components, as Lagrange functions and boundary matching are well established, but in their direct integration that bypasses the formal machinery of Bloch operators. This results in a method where the physical content (Schr\"{o}dinger equation at interior points, outgoing wave condition at the boundary) is transparently encoded in the matrix structure, facilitating both implementation and extension to more complex problems.

The approach is characterized by two essential features:

\textit{First}, I employ Lagrange-Legendre basis functions defined on a finite interval $[0, R]$. Unlike Lagrange-Laguerre functions, which contain an inherent exponential decay factor suited only for bound states, Lagrange-Legendre functions have no built-in asymptotic behavior. This neutrality allows them to represent both decaying and oscillatory functions within the computational domain, with the asymptotic character determined entirely by boundary conditions.

\textit{Second}, the outgoing wave boundary condition is imposed directly in the matrix equation. In R-matrix methods~\cite{Descouvemont2010}, the Bloch operator enforces a real boundary condition to maintain a symmetric matrix structure, and scattering amplitudes are extracted in a separate step by matching the R-matrix to asymptotic Coulomb functions. DBMM bypasses this two-step procedure by incorporating the complex logarithmic derivative $\psi'(R)/\psi(R) = \gamma_s$ directly in the last row of the matrix, where $\gamma_s = k H_\ell^{+\prime}/H_\ell^+$ evaluated at the boundary. This renders the matrix equation complex but yields scattering amplitudes in a single solve.

The relationship between DBMM and other bound-state techniques merits clarification. Unlike CSM, which transforms the asymptotic boundary condition through coordinate rotation, DBMM operates entirely in real coordinate space and requires explicit boundary matching at $r = R$. Like traditional R-matrix approaches, DBMM employs a finite computational domain with boundary matching, but eliminates the Bloch operator formalism.

The resulting method offers several practical advantages: (i) conceptual transparency, with each row of the matrix equation corresponding to a clear physical statement (Schr\"{o}dinger equation at interior points, boundary condition at the surface); (ii) implementation simplicity, requiring only standard linear algebra; (iii) natural extension to coupled channels without additional formal machinery; and (iv) pedagogical clarity that makes the method accessible for teaching nuclear reaction theory. While the computational cost is comparable to existing methods, scaling as $O(N^3)$ for the matrix solve, the primary advantage lies in the directness of the formulation rather than numerical efficiency.

The remainder of this paper is organized as follows. Section~\ref{sec:theory} develops the theoretical framework, presenting the scattering equations, Lagrange-Legendre basis functions, and the construction of the matrix equation with direct boundary matching. Section~\ref{sec:coupled} extends the formalism to coupled-channel problems. Section~\ref{sec:results} presents benchmark calculations validating the method against established approaches. Section~\ref{sec:conclusion} concludes this work.

\section{Theoretical Framework}
\label{sec:theory}

I consider two-body single-channel scattering described by the radial Schr\"{o}dinger equation with an optical potential. The method presented here employs Lagrange-Legendre basis functions on a finite interval, with the outgoing wave boundary condition incorporated directly into the matrix equation.

\subsection{Scattering equation}

For a given partial wave with orbital angular momentum $\ell$, the radial Schr\"{o}dinger equation reads
\begin{equation}
\left[-\frac{d^2}{dr^2} + \frac{\ell(\ell+1)}{r^2} + U(r) - k^2\right]\psi_\ell(r) = 0,
\label{eq:schrodinger}
\end{equation}
where $U(r) = 2\mu V(r)/\hbar^2$ is the reduced potential, $\mu$ is the reduced mass, and $k = \sqrt{2\mu E}/\hbar$ is the wave number at center-of-mass energy $E$. For charged-particle scattering, $U(r)$ includes both the nuclear optical potential and the Coulomb interaction, with the latter separated into long-range and short-range components:
\begin{equation}
U_C(r) = U_C^L(r) + U_C^S(r),
\end{equation}
where the long-range part is the point Coulomb interaction
\begin{equation}
U_C^L(r) = \frac{2\mu}{\hbar^2} \frac{Z_1 Z_2 e^2}{r},
\end{equation}
and the short-range part $U_C^S(r)$ accounts for the finite-size correction when treating nuclei as uniformly charged spheres. The total short-range potential is then
\begin{equation}
U_{\text{sr}}(r) = U_N(r) + U_C^S(r),
\end{equation}
where $U_N(r)$ is the nuclear optical potential. The regular Coulomb function $F_\ell(\eta, kr)$ satisfies
\begin{equation}
\left[-\frac{d^2}{dr^2} + \frac{\ell(\ell+1)}{r^2} + U_C^L(r) - k^2\right] F_\ell(\eta, kr) = 0.
\label{eq:coulomb}
\end{equation}

I decompose the total wave function into an incident wave and a scattered wave:
\begin{equation}
\psi_\ell(r) = e^{i\sigma_\ell}\left[F_\ell(\eta, kr) + \psi_\ell^{\text{sc}}(r)\right],
\label{eq:decomposition}
\end{equation}
where $\sigma_\ell$ is the Coulomb phase shift and $\eta = Z_1 Z_2 e^2 \mu / (\hbar^2 k)$ is the Sommerfeld parameter. Substituting into Eq.~(\ref{eq:schrodinger}) and using Eq.~(\ref{eq:coulomb}), the scattered wave satisfies
\begin{multline}
\left[-\frac{d^2}{dr^2} + \frac{\ell(\ell+1)}{r^2} + U(r) - k^2\right]\psi_\ell^{\text{sc}}(r) \\
= -U_{\text{sr}}(r) F_\ell(\eta, kr).
\label{eq:scattered}
\end{multline}

The boundary conditions are:
\begin{itemize}
\item At the origin: $\psi_\ell^{\text{sc}}(0) = 0$ (regularity).
\item At $r = R$: outgoing wave behavior, $\psi_\ell^{\text{sc}}(r) \propto H_\ell^+(\eta, kr)$.
\end{itemize}
The outgoing wave condition at the boundary $R$ can be expressed as
\begin{equation}
\frac{d\psi_\ell^{\text{sc}}}{dr}\bigg|_{r=R} = \gamma_s \, \psi_\ell^{\text{sc}}(R),
\label{eq:bc}
\end{equation}
where the logarithmic derivative of the outgoing wave is
\begin{equation}
\gamma_s = k \frac{H_\ell^{+\prime}(\eta, kR)}{H_\ell^+(\eta, kR)},
\label{eq:gamma_s}
\end{equation}
with $H_\ell^+ = G_\ell + i F_\ell$ being the outgoing Coulomb-Hankel function. Note that $\gamma_s$ is complex.

\subsection{Lagrange-Legendre basis functions}

I employ Lagrange functions based on shifted Legendre polynomials, defined on the interval $[0, R]$. The mesh points $\{r_j\}_{j=1}^N$ are obtained from the zeros of the shifted Legendre polynomial:
\begin{equation}
P_N(2x_j - 1) = 0, \quad r_j = R \cdot x_j,
\end{equation}
where $x_j \in (0, 1)$. The associated Gauss-Legendre weights are
\begin{equation}
\lambda_j = \frac{1}{4 x_j (1 - x_j) [P_N'(2x_j - 1)]^2}.
\end{equation}

The $x$-regularized Lagrange functions, which vanish at the origin, are defined as~\cite{Baye2015}
\begin{equation}
\hat{f}_j(x) = (-1)^{N-j} \sqrt{\frac{1-x_j}{x_j}} \frac{x \, P_N(2x-1)}{x - x_j},
\label{eq:lagrange_basis}
\end{equation}
where $x = r/R$. These functions satisfy the Lagrange condition at mesh points:
\begin{equation}
\hat{f}_j(x_i) = \frac{\delta_{ij}}{\sqrt{\lambda_j}}.
\end{equation}

The wave function is expanded as
\begin{equation}
\psi_\ell^{\text{sc}}(r) = R^{-1/2} \sum_{j=1}^{N} c_j \, \hat{f}_j(x),
\label{eq:expansion}
\end{equation}
where the expansion coefficients are related to function values by
\begin{equation}
c_j = \sqrt{R} \, \psi_\ell^{\text{sc}}(r_j) \sqrt{\lambda_j}.
\label{eq:coefficients}
\end{equation}

A key advantage of Lagrange-Legendre functions over Lagrange-Laguerre functions is their polynomial nature. Laguerre-based functions contain an inherent exponential decay factor, making them suitable for bound states or scattering calculations within the complex scaling method where wave functions decay asymptotically. Legendre-based functions impose no asymptotic behavior, allowing the boundary condition to fully determine whether the solution represents a bound or scattering state.

\subsection{Baye's derivative matrices}

The differential operators are represented by exact analytical matrices derived by Baye~\cite{Baye2015}. For the kinetic energy operator $T = -d^2/dr^2$ , the matrix elements in the coefficient representation are:
\begin{equation}
T_{ij} = \begin{cases}
\displaystyle (-1)^{i-j} \frac{x_i + x_j - 2x_i^2}{R^2 x_j(x_j - x_i)^2} \sqrt{\frac{x_j(1-x_j)}{x_i(1-x_i)^3}}, & i \neq j, \\[12pt]
\displaystyle \frac{N(N+1)x_i(1-x_i) - 3x_i + 1}{3R^2 x_i^2(1-x_i)^2}, & i = j.
\end{cases}
\label{eq:T_matrix}
\end{equation}

These matrices act on the expansion coefficients $\{c_j\}$, not directly on function values. This distinction is crucial: the Gauss quadrature properties of the Lagrange mesh ensure that these analytical formulas provide spectral accuracy for the differential operators.

\subsection{Matrix equation with direct boundary matching}

The discretized Schr\"{o}dinger equation at interior mesh points ($i = 1, 2, \ldots, N-1$) reads
\begin{equation}
\sum_{j=1}^{N} M_{ij} \, c_j = b_i,
\label{eq:interior}
\end{equation}
where the matrix elements are
\begin{equation}
M_{ij} = T_{ij} + \left[\frac{\ell(\ell+1)}{r_i^2} + U(r_i) - k^2\right] \delta_{ij},
\label{eq:M_interior}
\end{equation}
and the source term is
\begin{equation}
b_i = -U_{\text{sr}}(r_i) \, F_\ell(\eta, kr_i) \sqrt{R\lambda_i}.
\label{eq:source}
\end{equation}

The key innovation of DBMM is the treatment of the boundary condition. Instead of using a Bloch operator or a separate matching step, I directly impose the outgoing wave condition~(\ref{eq:bc}) in the last row of the matrix equation. Defining
\begin{equation}
B_j = \frac{d\hat{f}_j}{dx}\bigg|_{x=1} - R \gamma_s \, \hat{f}_j(1),
\label{eq:Bj}
\end{equation}
the boundary condition at $i = N$ becomes
\begin{equation}
\sum_{j=1}^{N} c_j B_j = 0.
\label{eq:bc_row}
\end{equation}

Using $P_N(1) = 1$ and $P_N'(1) = N(N+1)/2$, the basis functions at the boundary $x = 1$ have the analytical forms:
\begin{equation}
\hat{f}_j(1) = \frac{(-1)^{N-j}}{\sqrt{x_j(1-x_j)}},
\label{eq:fj_at_R}
\end{equation}
\begin{equation}
\frac{d\hat{f}_j}{dx}\bigg|_{x=1} = \frac{(-1)^{N-j}}{\sqrt{x_j(1-x_j)}} \left[N(N+1) - \frac{x_j}{1-x_j}\right].
\label{eq:fjprime_at_R}
\end{equation}

The complete $N \times N$ matrix equation takes the form
\begin{equation}
\begin{pmatrix}
M_{11} & \cdots & M_{1N} \\
\vdots & \ddots & \vdots \\
M_{N-1,1} & \cdots & M_{N-1,N} \\
B_1 & \cdots & B_N
\end{pmatrix}
\begin{pmatrix}
c_1 \\ \vdots \\ c_{N-1} \\ c_N
\end{pmatrix}
=
\begin{pmatrix}
b_1 \\ \vdots \\ b_{N-1} \\ 0
\end{pmatrix},
\label{eq:full_matrix}
\end{equation}
where the first $N-1$ rows correspond to the Schr\"{o}dinger equation at interior mesh points, and the last row encodes the boundary condition~(\ref{eq:Bj}). The matrix is complex (due to $\gamma_s$) and non-symmetric.

\subsection{Extraction of scattering observables}

After solving the linear system~(\ref{eq:full_matrix}) for the coefficients $\{c_j\}$, the scattered wave function at the boundary is
\begin{equation}
\psi_\ell^{\text{sc}}(R) = R^{-1/2} \sum_{j=1}^{N} c_j \, \hat{f}_j(1).
\end{equation}
The scattering amplitude is then
\begin{equation}
f_\ell = \frac{\psi_\ell^{\text{sc}}(R)}{k \, H_\ell^+(\eta, kR)},
\label{eq:amplitude}
\end{equation}
and the S-matrix element is
\begin{equation}
S_\ell = 1 + 2ik f_\ell.
\label{eq:smatrix}
\end{equation}
The phase shift $\delta_\ell$ (complex for absorptive potentials) follows from $S_\ell = e^{2i\delta_\ell}$, and the partial-wave reaction cross section is
\begin{equation}
\sigma_{\ell} = \frac{\pi}{k^2}(2\ell + 1)(1 - |S_\ell|^2).
\label{eq:cross_section}
\end{equation}

\section{Coupled-Channel Extension}
\label{sec:coupled}

The DBMM formalism extends naturally to coupled-channel problems, where multiple reaction channels are simultaneously active. The matrix structure of the method accommodates channel coupling without requiring additional formal machinery such as generalized Bloch operators.

For $n_c$ coupled channels labeled by $\alpha = 1, 2, \ldots, n_c$, with incident channel $\beta_0$, the wave function is decomposed as
\begin{equation}
\psi_{\alpha\beta_0}(r) = e^{i\sigma_{\ell_{\beta_0}}}\left[\delta_{\alpha\beta_0} F_{\ell_{\beta_0}}(\eta_{\beta_0}, k_{\beta_0} r) + \psi_{\alpha\beta_0}^{\text{sc}}(r)\right],
\end{equation}
where $\sigma_{\ell_{\beta_0}}$ is the Coulomb phase shift of the entrance channel and the incident wave $F_{\ell_{\beta_0}}$ appears only in the entrance channel. Substituting into the coupled-channel Schr\"{o}dinger equation yields the inhomogeneous equation for the scattered wave:
\begin{multline}
\left[-\frac{d^2}{dr^2} + \frac{\ell_\alpha(\ell_\alpha+1)}{r^2} + U_{\alpha\alpha}(r) - k_\alpha^2\right]\psi_{\alpha\beta_0}^{\text{sc}}(r) \\
+ \sum_{\beta \neq \alpha} U_{\alpha\beta}(r) \psi_{\beta\beta_0}^{\text{sc}}(r) = -U_{\alpha\beta_0}^{(\text{eff})}(r) F_{\ell_{\beta_0}}(\eta_{\beta_0}, k_{\beta_0} r),
\label{eq:coupled}
\end{multline}
where $\ell_\alpha$, $k_\alpha$, and $\eta_\alpha$ are the orbital angular momentum, wave number, and Sommerfeld parameter for channel $\alpha$, respectively. The effective source potential is
\begin{equation}
U_{\alpha\beta_0}^{(\text{eff})}(r) = \begin{cases}
U_{\beta_0\beta_0}^{\text{sr}}(r), & \alpha = \beta_0, \\
U_{\alpha\beta_0}(r), & \alpha \neq \beta_0.
\end{cases}
\end{equation}
For the entrance channel ($\alpha = \beta_0$), only the short-range part of the diagonal potential contributes because the Coulomb interaction is already accounted for in the regular Coulomb function $F_{\ell_{\beta_0}}$. For other channels, the full coupling potential $U_{\alpha\beta_0}$ (including any Coulomb multipole couplings) drives the transition.

Each scattered wave satisfies the outgoing boundary condition at $r = R$:
\begin{equation}
\frac{d\psi_{\alpha\beta_0}^{\text{sc}}}{dr}\bigg|_{r=R} = \gamma_{s,\alpha} \, \psi_{\alpha\beta_0}^{\text{sc}}(R),
\label{eq:cc_bc}
\end{equation}
with the channel-dependent logarithmic derivative
\begin{equation}
\gamma_{s,\alpha} = k_\alpha \frac{H_{\ell_\alpha}^{+\prime}(\eta_\alpha, k_\alpha R)}{H_{\ell_\alpha}^+(\eta_\alpha, k_\alpha R)}.
\end{equation}

Each scattered wave is expanded in the Lagrange-Legendre basis:
\begin{equation}
\psi_{\alpha\beta_0}^{\text{sc}}(r) = R^{-1/2} \sum_{j=1}^{N} c_j^{(\alpha\beta_0)} \hat{f}_j(x).
\end{equation}
The coupled-channel problem leads to a block matrix equation of dimension $n_c N \times n_c N$:
\begin{equation}
\begin{pmatrix}
M^{(11)} & \cdots & M^{(1n_c)} \\
\vdots & \ddots & \vdots \\
M^{(n_c 1)} & \cdots & M^{(n_c n_c)}
\end{pmatrix}
\begin{pmatrix}
\mathbf{c}^{(1\beta_0)} \\ \vdots \\ \mathbf{c}^{(n_c\beta_0)}
\end{pmatrix}
=
\begin{pmatrix}
\mathbf{b}^{(1)} \\ \vdots \\ \mathbf{b}^{(n_c)}
\end{pmatrix},
\label{eq:cc_matrix}
\end{equation}
where $\mathbf{c}^{(\alpha\beta_0)} = (c_1^{(\alpha\beta_0)}, \ldots, c_N^{(\alpha\beta_0)})^T$.

The diagonal block $M^{(\alpha\alpha)}$ has the same structure as the single-channel case. For interior rows ($i = 1, \ldots, N-1$):
\begin{equation}
M^{(\alpha\alpha)}_{ij} = T_{ij} + \left[\frac{\ell_\alpha(\ell_\alpha+1)}{r_i^2} + U_{\alpha\alpha}(r_i) - k_\alpha^2\right] \delta_{ij}.
\end{equation}
The last row encodes the channel-specific boundary condition:
\begin{equation}
M^{(\alpha\alpha)}_{Nj} = \frac{d\hat{f}_j}{dx}\bigg|_{x=1} - R \gamma_{s,\alpha} \, \hat{f}_j(1).
\end{equation}

The source vector for channel $\alpha$ depends on the incident channel $\beta_0$:
\begin{equation}
b_i^{(\alpha)} = \begin{cases}
-U_{\alpha\beta_0}^{(\text{eff})}(r_i) F_{\ell_{\beta_0}}(\eta_{\beta_0}, k_{\beta_0} r_i) \sqrt{R\lambda_i}, & i < N, \\
0, & i = N.
\end{cases}
\end{equation}

The off-diagonal blocks $M^{(\alpha\beta)}$ ($\alpha \neq \beta$) contain the coupling potentials at interior points, with the last row set to zero:
\begin{equation}
M^{(\alpha\beta)}_{ij} = \begin{cases}
U_{\alpha\beta}(r_i) \delta_{ij}, & i < N, \\
0, & i = N.
\end{cases}
\end{equation}

After solving the block system~(\ref{eq:cc_matrix}) for a given incident channel $\beta_0$, the S-matrix elements $S_{\alpha\beta_0}$ are obtained from the scattered wave at the boundary:
\begin{equation}
\psi_{\alpha\beta_0}^{\text{sc}}(R) = R^{-1/2} \sum_{j=1}^{N} c_j^{(\alpha\beta_0)} \hat{f}_j(1),
\end{equation}
\begin{equation}
f_{\alpha\beta_0} = \frac{\psi_{\alpha\beta_0}^{\text{sc}}(R)}{k_\alpha H_{\ell_\alpha}^+(\eta_\alpha, k_\alpha R)}, \quad
S_{\alpha\beta_0} = \delta_{\alpha\beta_0} + 2i \sqrt{\frac{k_\alpha}{k_{\beta_0}}} k_\alpha f_{\alpha\beta_0}.
\label{eq:cc_smatrix}
\end{equation}
The velocity factor $\sqrt{k_\alpha/k_{\beta_0}}$ ensures proper normalization of the S-matrix for channels with different wave numbers, arising from the flux conservation requirement. For elastic scattering ($\alpha = \beta_0$), this factor reduces to unity. The full S-matrix is constructed by repeating the calculation for each incident channel $\beta_0 = 1, 2, \ldots, n_c$.

\section{Results}
\label{sec:results}

To validate the DBMM approach, I present benchmark calculations for proton scattering on $^{12}$C at $E_{\rm lab} = 30$ MeV. The calculations employ the Koning-Delaroche global optical potential~\cite{Koning2003}, which provides a well-tested parameterization for nucleon-nucleus scattering. Results are compared against the established Numerov integration method~\cite{Thorlacius1987}, which serves as a reference for numerical accuracy.

The Koning-Delaroche potential includes real and imaginary volume terms, an imaginary surface term, and spin-orbit coupling. To simplify the calculation, I ignore the particle spin, which does not affect the conclusions. The potential parameters at $E_{\rm lab} = 30$ MeV are: real volume depth $V = 45.44$ MeV with $r_v = 1.127$ fm and $a_v = 0.676$ fm; imaginary volume depth $W = 3.08$ MeV with the same geometry; and imaginary surface depth $W_d = 6.37$ MeV with $r_d = 1.306$ fm and $a_d = 0.525$ fm. The Coulomb radius parameter is $r_c = 1.538$ fm.

The DBMM calculations use $N = 52$ Lagrange-Legendre mesh points on the interval $[0, R]$ with $R = 15$ fm, which achieves S-matrix accuracy better than $10^{-4}$ according to the convergence study below. The Numerov reference calculations employ a step size of $h = 0.02$ fm with matching performed at the same boundary radius. The calculations are performed using SLAM.jl~\cite{SLAM.jl}, an open-source Julia package developed by the author.

\begin{figure}[htbp]
\centering
\includegraphics[width=0.95\columnwidth]{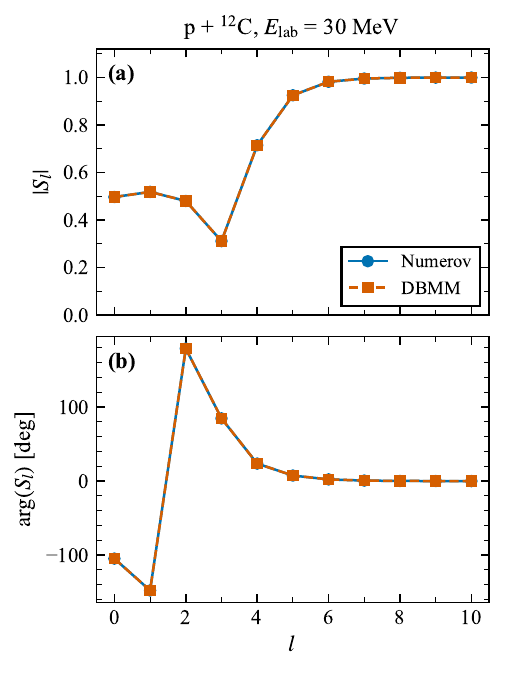}
\caption{S-matrix elements for p + $^{12}$C scattering at $E_{\rm lab} = 30$ MeV. (a) Modulus $|S_\ell|$ and (b) phase $\arg(S_\ell)$ as functions of orbital angular momentum $\ell$. Circles: Numerov method; squares: DBMM. The two methods show excellent agreement for all partial waves.}
\label{fig:smatrix}
\end{figure}

Figure~\ref{fig:smatrix} compares the S-matrix elements computed by DBMM (squares) and the Numerov method (circles) for partial waves $\ell = 0$ to $10$. Panel~(a) shows the modulus $|S_\ell|$, which quantifies the absorption strength: $|S_\ell| = 1$ corresponds to pure elastic scattering, while $|S_\ell| < 1$ indicates flux loss to inelastic channels. Low partial waves ($\ell \leq 3$) exhibit significant absorption with $|S_\ell| \approx 0.3$--$0.5$, reflecting their penetration into the nuclear interior where the imaginary potential is strongest. Higher partial waves progressively approach unity as the centrifugal barrier prevents nuclear overlap.

Panel~(b) displays the phase $\arg(S_\ell)$, which encodes the nuclear phase shift information beyond pure Coulomb scattering. The phase varies rapidly for low partial waves, changing from $-104^\circ$ at $\ell = 0$ through $-147^\circ$ at $\ell = 1$ to $+169^\circ$ at $\ell = 2$, indicating strong nuclear interaction effects. For $\ell \geq 6$, the phase approaches zero as these partial waves experience predominantly Coulomb scattering.

The agreement between DBMM and Numerov is excellent across all partial waves, with the two sets of symbols overlapping completely. Quantitatively, the differences in $|S_\ell|$ are below $2.5 \times 10^{-5}$, and the phase differences are less than $0.01^\circ$. This level of agreement demonstrates that the direct boundary matching approach achieves the same accuracy as traditional integration methods.

\begin{figure}[htbp]
\centering
\includegraphics[width=0.95\columnwidth]{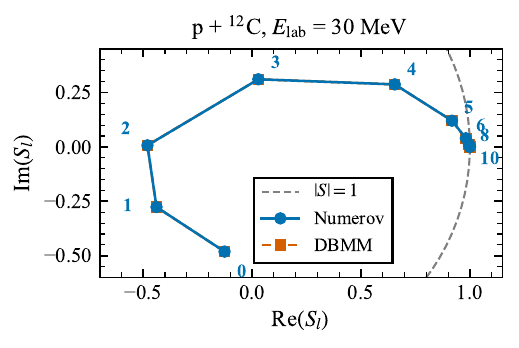}
\caption{Argand diagram showing the S-matrix trajectory in the complex plane for p + $^{12}$C at $E_{\rm lab} = 30$ MeV. Numbers indicate the orbital angular momentum $\ell$. The dashed gray circle represents the unitarity limit $|S| = 1$. Low partial waves lie inside the unit circle (absorption), while high partial waves approach unity (elastic scattering).}
\label{fig:argand}
\end{figure}

The S-matrix trajectory in the complex plane provides additional physical insight. Figure~\ref{fig:argand} shows the Argand diagram, where each point represents an S-matrix element $S_\ell = \Re(S_\ell) + i\Im(S_\ell)$. The dashed circle indicates the unitarity limit $|S| = 1$, which would be satisfied in the absence of absorption.

The trajectory begins at $\ell = 0$ in the lower half-plane with $|S_0| \approx 0.50$, well inside the unit circle due to strong absorption. As $\ell$ increases, the trajectory spirals counterclockwise, crossing into the upper half-plane at $\ell = 2$ and approaching the unit circle for high partial waves. By $\ell = 10$, the S-matrix element lies essentially on the unit circle, indicating negligible absorption.

This behavior reflects the classical interpretation of partial waves: low-$\ell$ waves correspond to impact parameters smaller than the nuclear radius and experience strong absorption, while high-$\ell$ waves pass outside the nuclear potential and scatter primarily via the long-range Coulomb interaction.

\begin{figure}[htbp]
\centering
\includegraphics[width=0.95\columnwidth]{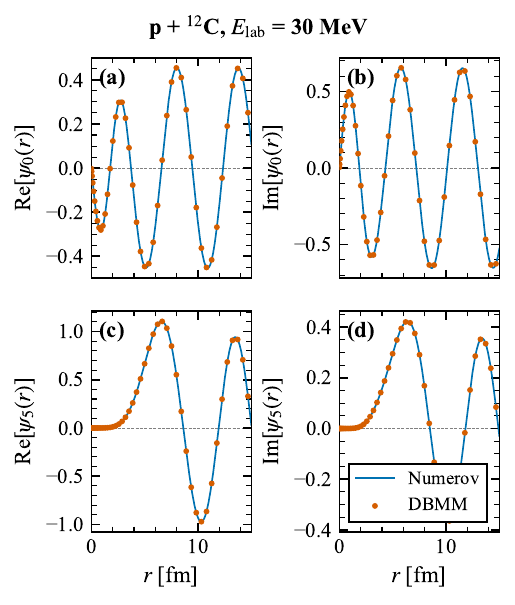}
\caption{Radial wave functions for p + $^{12}$C at $E_{\rm lab} = 30$ MeV. (a) Real part and (b) imaginary part for $\ell = 0$; (c) real part and (d) imaginary part for $\ell = 5$. Solid lines: Numerov method; circles: DBMM evaluated at Lagrange mesh points. The excellent agreement validates the Lagrange-Legendre expansion across all partial waves.}
\label{fig:wavefunction}
\end{figure}

Figure~\ref{fig:wavefunction} compares the radial wave functions computed by DBMM (circles) and the Numerov method (solid lines) for representative partial waves $\ell = 0$ and $5$. Panels~(a) and~(b) show the real and imaginary parts of the $\ell = 0$ wave function, respectively, while panels~(c) and~(d) display the corresponding results for $\ell = 5$.

For $\ell = 0$ [panels~(a) and~(b)], the wave function penetrates to small radii and exhibits oscillatory behavior throughout the interaction region. The real and imaginary parts have comparable magnitudes, reflecting the complex phase accumulated from the absorptive potential. For $\ell = 5$ [panels~(c) and~(d)], the wave function shows strong centrifugal suppression, with negligible amplitude inside $r = 4$ fm.

In all cases, the DBMM mesh points fall precisely on the Numerov curves, confirming that the Lagrange-Legendre expansion accurately captures the wave function throughout the computational domain. This agreement extends from the nuclear interior, where the potential is strongest, to the asymptotic region where the outgoing wave boundary condition is imposed.

It is instructive to compare the computational characteristics of DBMM with other methods. Traditional Numerov integration scales as $O(N)$ where $N$ is the number of integration steps. For instance, at $R = 10$~fm, Numerov with step size $h = 0.1$~fm requires $N \sim 100$ steps with $O(N)$ complexity, while DBMM requires $N \sim 32$ mesh points but with $O(N^3)$ complexity for the matrix solve. In fact, for single-channel problems, Numerov is faster than all matrix-based methods including both R-matrix and DBMM. The value of matrix-based methods lies elsewhere: in the natural extension to coupled channels, in the direct access to wave function coefficients for emulator construction, and in compatibility with other theoretical frameworks.

Both DBMM and R-matrix employ the same Lagrange-Legendre mesh and require solving a linear system of dimension $N \times N$ (or $n_c N \times n_c N$ for coupled channels), leading to $O(N^3)$ scaling for the matrix inversion. For R-matrix calculations, the propagation technique~\cite{LightWalker1976} can be employed to efficiently extend the solution to larger radii if needed, which is not directly applicable to DBMM.

To quantify the convergence behavior for the present p~+~$^{12}$C system, Table~\ref{tab:convergence} shows the minimum number of mesh points $N_{\min}$ required to achieve different phase shift accuracies for various boundary radii $R$. Here $\Delta\delta$ denotes the maximum phase shift difference over $\ell = 0$--$10$ with respect to a reference solution computed using Numerov integration with step size $h = 0.02$~fm.

\begin{table}[htbp]
\centering
\caption{Minimum number of Lagrange-Legendre mesh points $N_{\min}$ required to achieve different phase shift accuracies for p~+~$^{12}$C scattering at $E_{\rm lab} = 30$~MeV, as a function of boundary radius $R$.}
\label{tab:convergence}
\begin{tabular}{cccc}
\hline
$R$ (fm) & $|\Delta\delta| < 1^\circ$ & $|\Delta\delta| < 0.1^\circ$ & $|\Delta\delta| < 0.01^\circ$ \\
\hline
10 & 24 & 32 & 40 \\
12 & 26 & 32 & 40 \\
15 & 30 & 36 & 52 \\
20 & 36 & 48 & 64 \\
25 & 40 & 58 & 84 \\
30 & 48 & 62 & 80 \\
\hline
\end{tabular}
\end{table}

In the present p~+~$^{12}$C test, DBMM requires the $N_{\min}$ values reported in Table~\ref{tab:convergence}. Compared with the generally faster convergence often observed for the Bloch-operator (variational) R-matrix formulation, I estimate that DBMM may require $\alpha \sim 1.5$--$2$ times more mesh points to reach the same phase-shift accuracy. A plausible explanation is that the Bloch-operator formulation of R-matrix is closely connected to variational principles, for which scattering observables often converge faster than the wave function itself, whereas DBMM enforces the outgoing boundary condition directly, leading to a complex non-symmetric system without such variational protection. Since both methods are dominated by solving a dense linear system, the runtime scales as $T \propto N^3$ (LU factorization). If DBMM requires $N_{\text{DBMM}} = \alpha N_{\text{Rm}}$ mesh points to reach the same accuracy, the expected runtime ratio is
\begin{equation}
\frac{T_{\text{DBMM}}}{T_{\text{Rm}}} \approx \alpha^3.
\end{equation}
For $\alpha \simeq 1.5$--$2$, this gives $T_{\text{DBMM}}/T_{\text{Rm}} \approx 3$--$8$, depending on the specific problem. Absolute wall-clock times depend on hardware and implementation details.

Therefore, DBMM is \textit{not} expected to outperform the R-matrix method in raw computational speed for individual calculations, particularly when the propagation technique is employed. Nevertheless, DBMM offers two significant advantages that make it valuable in specific contexts:

\textit{First}, DBMM provides the scattered wave function directly as part of the solution. In contrast, R-matrix methods yield the R-matrix at the boundary, requiring an additional matching step to extract the S-matrix; if the interior wave function over the full radial interval is needed, additional reconstruction steps may be required depending on the specific R-matrix implementation. This direct access to wave functions is critical for reduced-basis emulator applications~\cite{Lei2025emulator}. In the reduced-basis approach, the solution vector contains wave function coefficients for all coupled channels at all mesh points, capturing the correlated behavior across channels simultaneously. This enables construction of a single unified reduced basis across all channels via singular value decomposition, avoiding the unfavorable $O(n_c^4)$ scaling of channel-wise approaches~\cite{CatacoraRios2024}. The resulting emulator achieves speedups of $10^3$--$10^4$ for continuum-discretized coupled-channels (CDCC)~\cite{Austern1987} calculations involving tens of coupled channels, with the reduced basis dimension ($n_b \sim 5$--$50$) governed by the smoothness of parameter dependence rather than the channel count. This massive acceleration, which I have already implemented and validated in Ref.~\cite{Lei2025emulator}, far outweighs the per-calculation overhead discussed above, making DBMM well suited for emulator construction.

\textit{Second}, DBMM can serve as an alternative boundary-matching scheme within the complex scaling framework for few-body scattering~\cite{Lazauskas2011,Lazauskas2012a,Lazauskas2012b,Lazauskas2019}. The complex scaling method, which rotates coordinates as $r \to r e^{i\theta}$ to convert oscillatory scattering states into exponentially decaying functions, is powerful for treating few-body scattering with complex boundary conditions in Faddeev-type equations. However, at low energies, the small wave number $k$ requires large rotation angles $\theta$ to ensure adequate damping [the wave function decays as $\exp(-k \sin\theta \cdot r)$], which can introduce numerical difficulties in potential evaluation. By providing an alternative real-coordinate approach that can be interfaced with complex-scaled interior solutions, DBMM offers a pathway to extend few-body scattering calculations to lower energies while maintaining the same code infrastructure.

\section{Conclusion}
\label{sec:conclusion}

I have presented a direct boundary matching method (DBMM) for solving nuclear scattering problems using Lagrange-Legendre basis functions. The method belongs to the family of bound-state techniques for the continuum, reformulating scattering problems into a form solvable within a localized $L^2$ basis.

The key feature of DBMM is the direct incorporation of the outgoing wave boundary condition into the last row of the matrix equation. This approach eliminates the need for Bloch operators and two-step matching procedures required in traditional R-matrix methods. Unlike the complex scaling method, which rotates coordinates into the complex plane to convert oscillatory scattering states into $L^2$-integrable functions, DBMM operates entirely in real coordinate space and yields scattering amplitudes through explicit boundary matching. The resulting matrix equation, though complex and non-symmetric, can be solved with standard linear algebra routines.

The formalism has been extended to coupled-channel problems. The wave function decomposition $\psi_{\alpha\beta_0} = \delta_{\alpha\beta_0} F_{\ell_{\beta_0}} + \psi_{\alpha\beta_0}^{\text{sc}}$ naturally leads to an effective source potential that distinguishes between the entrance channel (short-range potential only) and other channels (full coupling potential including Coulomb multipoles). The block matrix structure accommodates channel coupling without additional formal machinery.

Benchmark calculations for p~+~$^{12}$C scattering at $E_{\rm lab} = 30$~MeV demonstrate excellent agreement with the established Numerov integration method. The S-matrix elements agree to better than $2.5 \times 10^{-5}$ in modulus and $0.01^\circ$ in phase across all partial waves. The wave functions computed by DBMM match the Numerov solutions throughout the computational domain, from the nuclear interior to the asymptotic region.

The method offers several practical advantages: conceptual transparency, where each row of the matrix equation corresponds to a clear physical statement; implementation simplicity, requiring only standard linear algebra; and natural extension to coupled channels. While the computational cost scales as $O(N^3)$ for the matrix inversion, comparable to other methods, the primary contribution is methodological simplification rather than numerical efficiency. The directness of the formulation, encoding the Schr\"{o}dinger equation and boundary conditions in a single matrix equation, provides both pedagogical value and practical convenience for implementation. The open-source implementation in SLAM.jl~\cite{SLAM.jl} provides a readily available tool for nuclear reaction calculations.

Future directions include combining the method with microscopic structure calculations to provide a unified description of nuclear structure and reactions.

\begin{acknowledgments}
This work was supported by the National Natural Science Foundation of China (Grant Nos.~12475132 and 12535009) and the Fundamental Research Funds for the Central Universities.
\end{acknowledgments}

\bibliography{references}

\end{document}